\documentclass[conference]{IEEEtran}
\IEEEoverridecommandlockouts
\usepackage{cite}
\usepackage{amsmath,amssymb,amsfonts}
\usepackage{algorithmic}
\usepackage{graphicx}
\usepackage{textcomp}
\usepackage{xcolor}
\usepackage[caption=false,font=normalsize,labelfont=sf,textfont=sf]{subfig}
\usepackage{algorithm}
\usepackage{array}
\usepackage{textcomp}
\usepackage{stfloats}
\usepackage{xurl}
\usepackage{verbatim}
\usepackage{romannum}
\usepackage{makecell}
\usepackage{multirow}
\usepackage{hyperref}
\usepackage{booktabs}

\def\BibTeX{{\rm B\kern-.05em{\sc i\kern-.025em b}\kern-.08em
    T\kern-.1667em\lower.7ex\hbox{E}\kern-.125emX}}
\begin{document}

\title{Regime-Adaptive Weighted Ensemble Learning for Computing-Driven Dynamic Load Forecasting \\in AI Data Centers\\
\thanks{This research was supported by the U.S. National Science Foundation through award number 2418359 and the Hamm Institute of American Energy.

*Corresponding author: Ying Zhang}
}

\author{
 \IEEEauthorblockN{Ziying Wang, Ying Zhang*, Lei Wang}
    \IEEEauthorblockA{\textit{School of Electrical and Computer Engineering} \\
    \textit{Oklahoma State University}\\
    Stillwater, OK, U.S. \\
    \{zy.wang;y.zhang;leiwang\}@okstate.edu}\\

\and

\IEEEauthorblockN{Yuzhang Lin}
\IEEEauthorblockA{\textit{Department of Electrical and Computer Engineering} \\
\textit{New York University} \\
Brooklyn, NY, U.S.  \\
yuzhang.lin@nyu.edu}
}

\maketitle

\begin{abstract}
Short-term load forecasting for AI data centers presents new challenges because it is computing-driven, with heterogeneous job arrivals, sizes, and durations exhibiting bursty, non-stationary dynamics. Compared with traditional load types, data center loads are less researched and can pose greater threats to the efficiency and stability of power grids. To close the gap, this paper proposes a regime-adaptive ensemble learning forecasting algorithm to predict computing-driven dynamic workloads in AI data centers. A weight-learned neural network within an ensemble learning framework is developed to exploit the complementary strengths of two machine learning (ML) submodels across varying operating regimes. Furthermore, a novel feature engineering strategy is developed to incrementally learn from a non-stationary data stream. Thus, the ensemble weights are dynamically optimized to facilitate adaptive calibration of inter-submodel contributions. Comparative case studies on the MIT Supercloud dataset demonstrate that the proposed method significantly enhances load forecasting accuracy and adaptivity across various regimes, and the selected combination of ML models for ensemble learning outperforms other possible combinations. To the best of our knowledge, our method is the first to reduce minute-class forecasting errors for AI data center loads to below 1\%, highlighting its potential for grid-interactive coordination and demand response.
\end{abstract}


\section{Introduction}

Beyond the rapid growth in electricity demand, AI data centers differ fundamentally from traditional data centers in hosting denser, more data-intensive, and less predictable workloads, thereby introducing new operational risks to power system stability \cite{mishra2025understanding}. In particular, AI training jobs exhibit bursty power surges followed by short cool-down periods \cite{biswas2025evaluating}, with power demand rapidly escalating from idle (around 10\%) to full capacity (100–150\%) almost instantaneously \cite{quint2025assessment}. Aggregated across many such workloads, these dynamics yield continuous but highly non-stationary power variations that traverse multiple operating regimes at the AI data center level. The resulting fluctuations can cause frequency and voltage stability problems and induce persistent forced oscillations in the grid \cite{ko2025wide}. These characteristics highlight the need for dynamic AI data center load forecasting models that can support power system operation, optimization, and control. Among these, short-term load forecasting is crucial for power dispatch and preventive control under rapid AI workload fluctuations \cite{hu2021characterization, wang2024minute}.

Existing short-term load forecasting methods can be broadly classified into statistical and machine learning (ML) approaches. Early studies mainly adopted statistical models, such as moving-average and autoregressive families including autoregressive integrated moving average (ARIMA) \cite{hsu2018self}, because of their simple structure, low computational complexity, and ease of implementation. However, such models rely on linearity and stationarity assumptions, which limit their ability to capture the complex temporal dependence and rapidly changing operating regimes of AI data centers.
To this end, ML models with greater representational capacity have been explored for electric load prediction, including tree-based models such as extreme gradient boosting (XGBoost) \cite{kathiravan2023novel} and deep learning architectures such as long short-term memory (LSTM) and gated recurrent unit (GRU)\cite{chung2023artificial}. 

In the AI data center context, a recent study \cite{mughees2025short} investigates the short-term electric load forecasting accuracy of existing LSTM, GRU, and one-dimensional convolutional neural network (1D-CNN) models. Although these methods in \cite{mughees2025short} can better represent nonlinear load dynamics, their predictions fail to capture sharp power transients, including sudden dips and peaks, within minute-scale windows. 
These results suggest that an individual ML model with a single architecture cannot handle heterogeneous profiles across computing regimes and reflect an unsolved difficulty in capturing AI workload dynamics. 
The highly dynamic and non-cyclic power patterns of AI workloads \cite{chen2025electricity} differ substantially from traditional demand profiles and call for forecasting methods tailored to AI data center operation.
Ensemble learning offers a promising approach to improving forecast accuracy over individual ML predictors \cite{wu2017data, liao2019research}. 
However, most ensemble forecasting methods are originally designed for conventional load types and rely on fixed or globally optimized weights. These predictors may fail to respond to local rapid load ramps in AI data centers. This gap motivates the development of adaptive ensemble learning methods for AI data center load prediction that can adapt to all regimes without retraining and regime labeling.

\begin{figure*}[!t]
  \centering
  \includegraphics[width=0.99\textwidth]{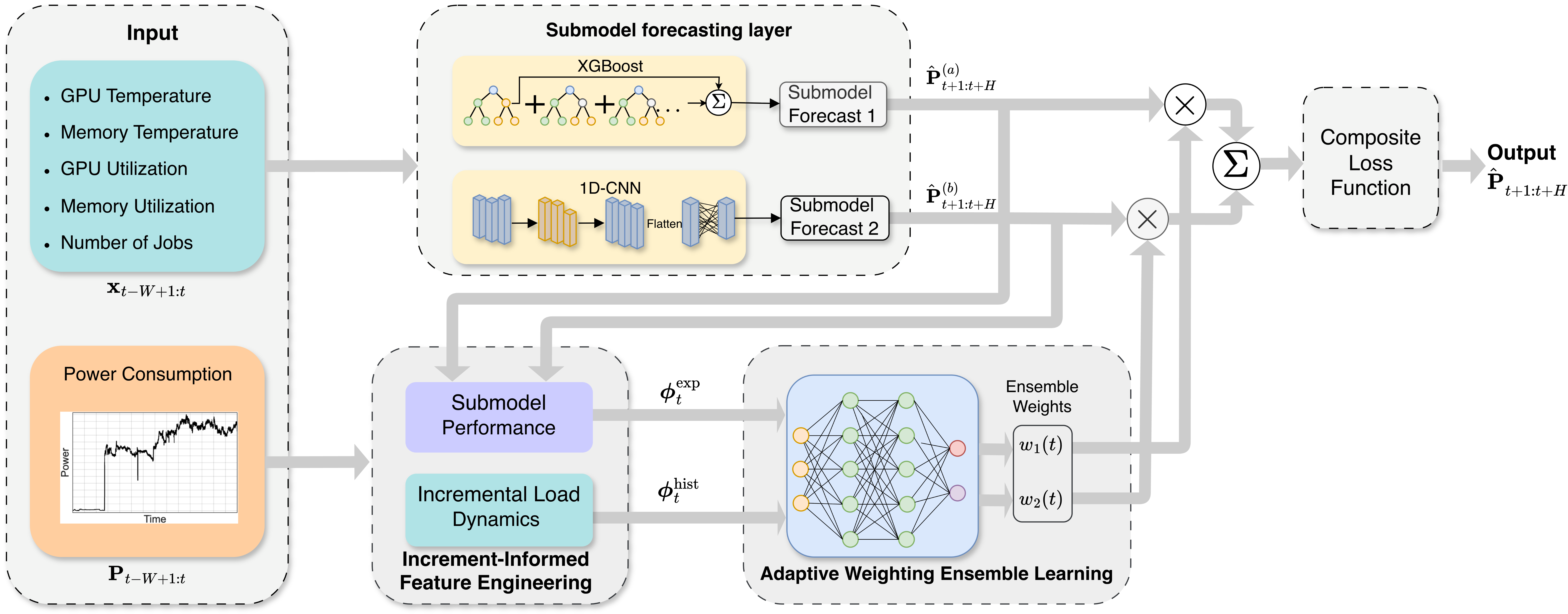}
  \caption{Architecture of the proposed regime-adaptive ensemble learning method that can adapt to various computing regimes.}
   \label{fig:algorithm}
\end{figure*}

To close the gaps, a regime-adaptive ensemble forecasting algorithm is proposed for short-term load forecasting of AI data centers. The proposed algorithm integrates two complementary submodels through an adaptive ensemble learning neural network, in which XGBoost captures cross-feature interactions, whereas the 1D-CNN applies convolutional filters to extract local temporal patterns. Furthermore, the ensemble weights are cast as extra learnable parameters of the assembled submodels to facilitate adaptive calibration of inter-submodel contributions during training.
Besides, dedicated increment-informed feature engineering is proposed to incrementally learn new information from a non-stationary data stream. The proposed ensemble thus adaptively adjusts its weights based on historical submodel performance and load dynamics. The main contributions are summarized as follows:
\begin{itemize}
    \item By integrating dedicated increment-informed feature engineering to adaptive weighting ensemble learning, the proposed method significantly improves short-term load forecasting accuracy across complex non-stationary AI data center operating regimes while exploiting the complementary predictive characteristics of individual ML models. Our method can anticipate ramping workload consumption due to abrupt job submissions and switches.
    \item Comparative experiments on the MIT Supercloud dataset show that the proposed method improves accuracy by about 80\% while adapting robustly to varying operating regimes, and largely outperforms recent ML-based predictors and their ensembles. To the best of our knowledge, our method is the first to reduce minute-class forecasting errors of AI data center loads to below 1\%.
\end{itemize}

\section{Proposed Regime-Adaptive Ensemble Learning forecasting Method}\label{sec:Methodology}
A regime-adaptive ensemble learning method is proposed for short-term AI data center load forecasting to achieve workload prediction across representative operating regimes, namely idle, ramp-up, high-demand, and ramp-down, which exhibit distinct dynamics. As shown in Fig.~\ref{fig:algorithm}, the proposed algorithm comprises submodel forecasting, feature engineering, and adaptive-weighting ensemble learning.

Submodel forecasts from XGBoost and 1D-CNN are generated using exogenous features. The two submodels exhibit complementary behavior across distinct operating regimes, a characteristic that can be exploited by ensemble learning to integrate their individual strengths. Subsequently, the historical power consumption and submodel forecasts are fed into the increment-informed feature construction stage, where the historical submodel performance and increment load dynamics are computed. These features are passed to the adaptive weighting ensemble learning stage to estimate time-varying ensemble weights, which are then used to aggregate the two submodel forecasts into the ensemble forecast. The proposed method fully exploits the complementary strengths of the two submodels under different operating regimes, thereby improving forecasting accuracy and adaptiveness.

\subsection{Forecasting Submodels: XGBoost and 1D-CNN}

Two top-performing forecasting submodels for AI data center loads, XGBoost and 1D-CNN, are developed to learn from recent operating regimes to predict future power demand. These models are preferred over more complex neural architectures, offering comparable accuracy at lower data and computational costs, which are critical for short-term forecasting in AI data center operations. In the case study, the complementarity of the selected submodels is evaluated and compared with ensembles of other ML models.

Let $P_t$ denote the cluster-level power consumption of an AI data center at time step $t$, and let $\mathbf{x}_t$ denote the corresponding exogenous feature vector, which characterizes the temporal condition, thermal status, resource utilization, and workload intensity. Given a historical observation window of length $W$, define the historical power sequence as $\mathbf{P}_{t-W+1:t}=\{P_{t-W+1}, P_{t-W+2}, \ldots, P_t\}$, and the aligned exogenous feature sequence as
$\mathbf{x}_{t-W+1:t}=\{\mathbf{x}_{t-W+1}, \mathbf{x}_{t-W+2}, \ldots, \mathbf{x}_t\}.$

In AI data centers, short-term power demand is driven by rapidly changing workloads, resource contention, and operating-regime transitions, rendering the load evolution highly nonlinear and non-stationary. The forecasting task is to predict the future power trajectory over a horizon of $H$ steps:
\begin{equation}
\hat{\mathbf{P}}_{t+1:t+H}
=f\!\left(\mathbf{P}_{t-W+1:t},\,\mathbf{x}_{t-W+1:t}\right).
\label{eq:general_forecast}
\end{equation}

XGBoost and 1D-CNN are selected as the two submodels based on their complementary expertise in different operating regimes. Alternative combinations are implemented in the case study to assess pairwise complementarity prior to comparing the adaptive ensemble forecasting performance.

\subsubsection{Extreme Gradient Boosting}
XGBoost expresses the predictor as an additive combination of regression trees, where each tree captures cross-feature interactions through hierarchical splits. For the $h$ step prediction, the XGBoost model is written as
\begin{equation}
\hat{y}_{i,h} = \sum_{k=1}^{K_h} f_{k,h}(\mathbf{z}_{i}), 
\qquad f_{k,h} \in \mathcal{F}, \quad h=1,2,\ldots,H
\end{equation}
where $\mathbf{z}_{i}$ is the input vector for sample $i$, $\hat{y}_{i,h}$ is the prediction at horizon $h$, $f_{k,h}(\cdot)$ denotes the $k$th regression tree of the model for step $h$, $K_h$ is the number of trees, and $\mathcal{F}$ is the space of regression trees. The multi-step prediction is obtained by combining the outputs of the $H$ step-specific regressors.

\subsubsection{One-Dimensional Convolutional Neural Network}
1D-CNN extracts temporal patterns by applying convolutional kernels along the time axis. For the $k$th filter in layer $\ell$, the convolution operation can be written as
\begin{equation}
z_{k}^{(\ell)}(\tau)
=
\psi
\left(
\sum_{c=1}^{C_{\ell-1}}
\sum_{r=0}^{R-1}
w_{k,c,r}^{(\ell)}
\, z_{c}^{(\ell-1)}(\tau-r)
+
b_{k}^{(\ell)}
\right)
\end{equation}
where $R$ is the kernel size, $C_{\ell-1}$ is the number of input channels, $w_{k,c,r}^{(\ell)}$ denotes the convolution weight, and $\psi(\cdot)$ is the activation function. By stacking multiple convolutional layers, 1D-CNN can effectively capture local short-term temporal variation patterns in the load trajectory.

These two submodels are trained by minimizing the standard mean squared errors (MSE) on the same training dataset.

\subsection{Proposed Increment-Informed Feature Engineering}

AI data center power demand exhibits unique computing-driven dynamics that distinguish it from conventional residential load or renewable generation profiles, motivating dedicated feature engineering. Thus, a novel feature engineering strategy is proposed to incrementally learn new information from a non-stationary stream of data for AI data center short-term load forecasting.

Two feature vectors, $\boldsymbol{\phi}^{\mathrm{hist}}_t$ and $\boldsymbol{\phi}^{\mathrm{exp}}_t$, are constructed as input to the adaptive ensemble weighting module, characterizing the recent operating regime and the divergence between the two submodels, respectively.

Define the incremental change over two successive time steps $t-1$ and $t$ as $\Delta P_{t} = P_{t} - P_{t-1}$. Based on the historical power sequence, the incremental load dynamics feature vector at time $t$ is constructed as
\begin{subequations}
\renewcommand{\theequation}{\theparentequation}
\begin{equation}
\boldsymbol{\phi}^{\mathrm{hist}}_t=\left\{P_t,\;|\Delta P_t|,\;\mu_{|\Delta P|},\;\sigma_{\Delta P},\;s_t\right\}
\end{equation}
\setcounter{equation}{0}
\renewcommand{\theequation}{\theparentequation\alph{equation}}
\begin{equation}
\mu_{|\Delta P|}=\frac{1}{W-1}\sum_{\tau=t-W+2}^{t}|\Delta P_{\tau}|
\end{equation}
\begin{equation}
\sigma_{\Delta P}=\sqrt{\frac{1}{W-1}\sum_{\tau=t-W+2}^{t}\left(\Delta P_{\tau} - \overline{\Delta P}\right)^2}
\end{equation}
\begin{equation}
s_t=\frac{P_t - P_{t-W+1}}{W-1}.
\end{equation}
\end{subequations}
where $\mu_{|\Delta P|}$ denotes the mean absolute increment over the historical window, and $\sigma_{\Delta P}$ denotes the standard deviation of the increments;
$\overline{\Delta P}=\frac{1}{W-1}\sum_{\tau=t-W+2}^{t}\Delta P_{\tau}$ is the mean increment, and $s_t$ is the average slope over the window. These features collectively characterize the recent operating regime of the AI data center through its power level, variability, and short-term trend, thereby providing contextual information for adaptive ensemble weighting.

Denote the two submodels' forecast outputs as
$\hat{\mathbf{P}}^{(a)}_{t+1:t+H}$ and $\hat{\mathbf{P}}^{(b)}_{t+1:t+H}$, respectively. The historical submodel-performance feature vector is constructed from the one-step-ahead predictions $\hat{P}^{(a)}_{t+1}$ and $\hat{P}^{(b)}_{t+1}$, as these are the most informative for characterizing the recent inter-submodel relationship:
\begin{subequations}
\renewcommand{\theequation}{\theparentequation}
\begin{equation}
\boldsymbol{\phi}^{\mathrm{exp}}_t
=
\left\{
\hat{P}^{(a)}_{t+1},\;
\hat{P}^{(b)}_{t+1},\;
d_t,\;
|d_t|,\;
r_t,\;
\Delta\hat{P}^{(a)}_{t+1},\;
\Delta{P}^{(b)}_{t+1}
\right\}
\label{fe_ex}
\end{equation}
\setcounter{equation}{0}
\renewcommand{\theequation}{\theparentequation\alph{equation}}
\begin{equation}
r_t=\frac{|d_t|}
{\frac{1}{W}\sum_{\tau=t-W+1}^{t}|P_{\tau}|}
\end{equation}
\end{subequations}
where $d_t = \hat{P}^{(a)}_{t+1} - \hat{P}^{(b)}_{t+1}$ denotes the signed divergence between the two one-step-ahead predictions, and $r_t$ denotes the divergence normalized by the recent power level;
$\Delta \hat{P}^{(a)}_{t+1} = \hat{P}^{(a)}_{t+1} - P_t$, and $\Delta \hat{P}^{(b)}_{t+1} = \hat{P}^{(b)}_{t+1} - P_t$ are the one-step-ahead predicted power increments relative to the current power level for the two submodels. These features reflect the consistency and divergence between the two submodels, as well as their deviations from the current power level, thereby providing informative cues for adaptive ensemble weighting.

The two feature vectors $\boldsymbol{\phi}^{\mathrm{hist}}_t$ and $\boldsymbol{\phi}^{\mathrm{exp}}_t$ jointly serve as the input to the proposed adaptive weighting ensemble learning network.

\subsection{Proposed Adaptive Weighting Ensemble Learning}

An adaptive weighting ensemble learning method is proposed to exploit the complementary predictive characteristics of individual ML models to improve short-term load forecasting performance under complex non-stationary AI data center operating regimes. The proposed method can anticipate dynamic workload consumption due to abrupt job submissions and switches.

Based on the constructed feature vectors, a three-layer MLP is employed to generate the ensemble weights adaptively:
\begin{equation}
\mathbf{h}^{(1)}_t
=
\mathrm{ReLU}
\left(
\mathbf{W}_1 [\boldsymbol{\phi}^{\mathrm{hist}}_t, \boldsymbol{\phi}^{\mathrm{exp}}_t] + \mathbf{b}_1
\right)\label{eq:8}
\end{equation}
\begin{equation}
\mathbf{h}^{(2)}_t
=
\mathrm{ReLU}
\left(
\mathbf{W}_2 {\mathbf{h}}^{(1)}_t + \mathbf{b}_2
\right)
\end{equation}
\begin{equation}
[w_1(t),w_2(t)]=\sigma\left(\mathbf{W}_{3}^{\top}\mathbf{h}^{(2)}_t +\mathbf{b_3}\right)\label{eq:10}
\end{equation}
where $\mathbf{W}_1$, $\mathbf{W}_2$, $\mathbf{W}_3$, $\mathbf{b}_1$, $\mathbf{b}_2$, and $\mathbf{b}_3$ are learnable parameters, and the parameter set is denoted as $\{\mathbf W,\mathbf b\}$; $\sigma(\cdot)$ denotes the softmax function that converts a vector of raw scores into a probability distribution, where each value is between 0 and 1 and the total sum equals 1; $w_1(t)$ and $w_2(t)$ denotes the ensemble weights for the submodels, respectively.

The ensemble weights are learned to dynamically adjust the contribution of each submodel via \eqref{eq:8}-\eqref{eq:10}. The weighting ensemble forecasted power is expressed through the learnable ensemble weights as
\begin{equation}
\hat{\mathbf{P}}^{\mathrm{ens}}_{t+1:t+H}=w_1(t)\hat{\mathbf{P}}^{(a)}_{t+1:t+H}+ w_2(t)\hat{\mathbf{P}}^{(b)}_{t+1:t+H}
\label{eq:ensemble_output}
\end{equation}

The primary training objective of the weight estimation network is to minimize the prediction loss between the ensemble output and the target trajectory: 
\begin{equation}
\mathcal{L}_{\mathrm{pred}}
=
\frac{1}{N}
\sum_{t=1}^{N}
\left(
\hat{P}^{\mathrm{ens}}_{t+1} - P_{t+1}
\right)^2.
\label{eq:pred_loss_h1}
\end{equation}
where $N$ denotes the number of training samples.

To avoid degenerate weighting, an interpolation coefficient is introduced. When the ground truth falls between the two submodel predictions, the coefficient is defined as
\begin{equation}
w^{\star}(t)=\frac{P_{t+1} - \hat{P}^{(b)}_{t+1}}{\hat{P}^{(a)}_{t+1} - \hat{P}^{(b)}_{t+1}}
\label{eq:ideal_w_raw}
\end{equation}

To supervise the ensemble weight $w_1(t)$ toward the interpolation coefficient, the auxiliary loss is defined as 
\begin{equation}
\mathcal{L}_{w}
=
\frac{1}{|\mathcal{V}|}
\sum_{t \in \mathcal{V}}
\left( w_1(t) - w^{\star}(t) \right)^2
\label{eq:weight_aux_loss}
\end{equation}
where $\mathcal{V}$ denotes the index set of samples for which the ground truth $P_{t+1}$ lies within the interval bounded by $\hat{P}^{(a)}_{t+1}$ and $\hat{P}^{(b)}_{t+1}$. 

The training objective of the proposed ensemble model is to minimize the following composite loss function:
\begin{equation}
\mathcal{L}_{\mathrm{ens}}
=
\mathcal{L}_{\mathrm{pred}}
+
\lambda \mathcal{L}_{w}
\label{eq:ensemble_loss}
\end{equation}
where $\lambda \geq 0$ controls the contribution of the auxiliary loss.

Utilizing the separately trained ML submodels, the proposed ensemble learning model is trained offline on historical data samples using the Adam optimizer with mini-batch stochastic gradient descent to optimize the parameters $\{w_1,w_2,\mathbf W,\mathbf b\}$. Once well-trained, the proposed method is deployed for online inference.

\section{Case Studies}\label{sec:CaseStudies}
To evaluate the proposed method, the MIT Supercloud dataset \cite{samsi2021supercloud} is adopted, which comprises job-level GPU logs at 0.1\,s resolution, including utilization, temperature, anonymized job identifiers, and per-GPU power. The raw job-level records are resampled into 1-min node-level series and aggregated across computing nodes to construct a cluster-level dataset. 
Mean GPU/memory temperatures and utilizations are computed across all nodes, and workload intensity is quantified by the number of active jobs per minute. As idle intervals are missing from the raw logs, each node is reindexed to a complete 1-min grid, with empty slots assigned fixed GPU baselines: $P_{\mathrm{idle}}=25\,\mathrm{W}$, zero utilization, $T_{\mathrm{idle}}^{(\mathrm{gpu})}=25.34^\circ\mathrm{C}$, and $T_{\mathrm{idle}}^{(\mathrm{mem})}=24.0^\circ\mathrm{C}$. The dataset is divided chronologically: 60\% for submodel training, 20\% for ensemble training, and the remaining for testing. All variables are normalized using a StandardScaler fitted on the submodel-training data.

The proposed ensemble learning method is compared with LSTM, SVR, XGBoost, and 1D-CNN \cite{kathiravan2023novel,mughees2025short}. Model accuracy is evaluated using the normalized root mean square error (NRMSE) and normalized mean absolute error (NMAE) \cite{song2025self}:
\begin{equation}
    NRMSE = \sqrt{\frac{1}{N_t}\sum_{t=1}^{N_t}\left(\frac{\hat{P}_t - P_t}{P_{\max}}\right)^2}
\end{equation}
\begin{equation}
    NMAE =  {\frac{1}{N_t}\sum_{t=1}^{N_t}\frac{|\hat{P}_t - P_t|}{P_{\max}}}
\end{equation}
where $\hat{P}_t$ and $P_t$ denote the actual and predicted power at time step $t$, and $N_t$ is the number of test samples; $P_{\max}$ is the maximum value of the acutal power.
In addition, a comparative analysis of combinations of these ML submodels, selected based on recent research on AI data center load forecasting \cite{mughees2025short}, is conducted using the same metrics.

All experiments are conducted on a workstation with an Intel Xeon processor at 2.20\,GHz and an NVIDIA L4 GPU with 23\,GB memory.

\begin{table}[!t]
\captionsetup{font=small,justification=centering}
\caption{Comparison of Forecasting Errors with Other ML methods}
\centering
\label{tab:one_step_error_comparison}
\begin{tabular}{|l|c|c|}
\hline
\textbf{Method} & \textbf{NRMSE [\%]} & \textbf{NMAE [\%]} \\
\hline
SVR   & 3.04 & 2.30 \\\hline
LSTM   & 2.66 & 2.11 \\\hline
XGBoost& 2.46 & 1.86  \\
\hline
1D-CNN  & 2.45 & 1.85 \\
\hline
\textbf{Proposed Method} & \textbf{0.83} & \textbf{0.37} \\
\hline
\end{tabular}%
\end{table}

\begin{figure}[!t]
  \centering
  \vspace{-5pt}
  \subfloat[]{%
    \includegraphics[width=0.46\textwidth]{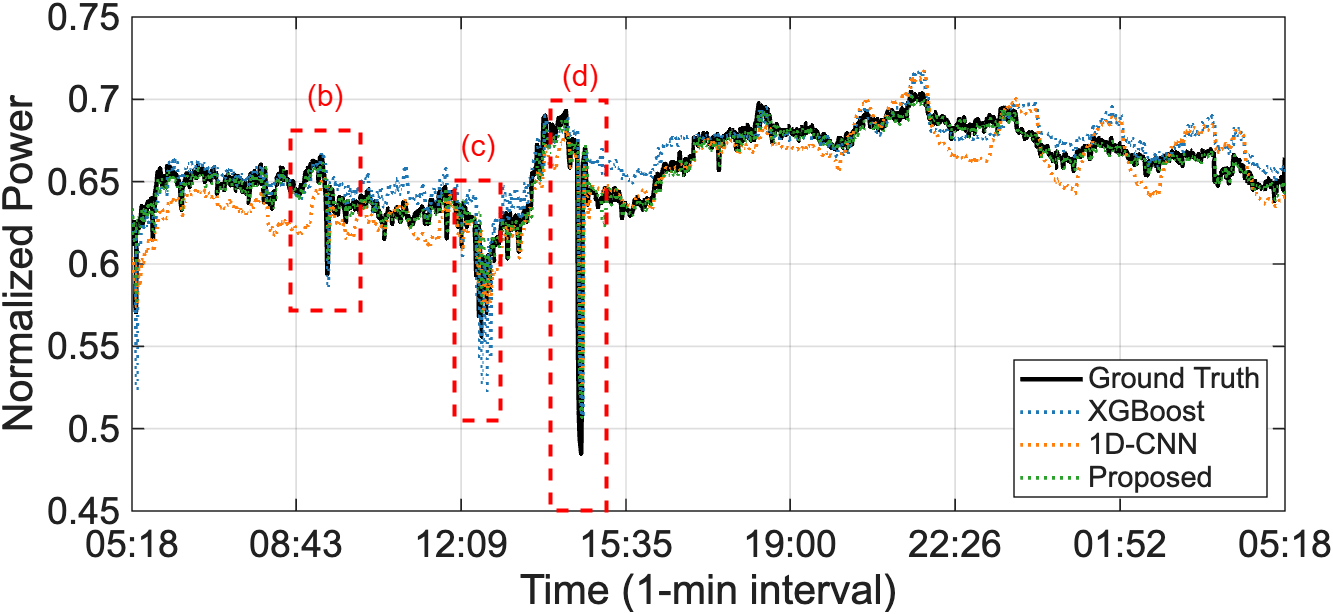}%
    \label{fig:Figure_1}}
    \\[4pt]
  \subfloat[]{%
    \includegraphics[width=0.40\textwidth]{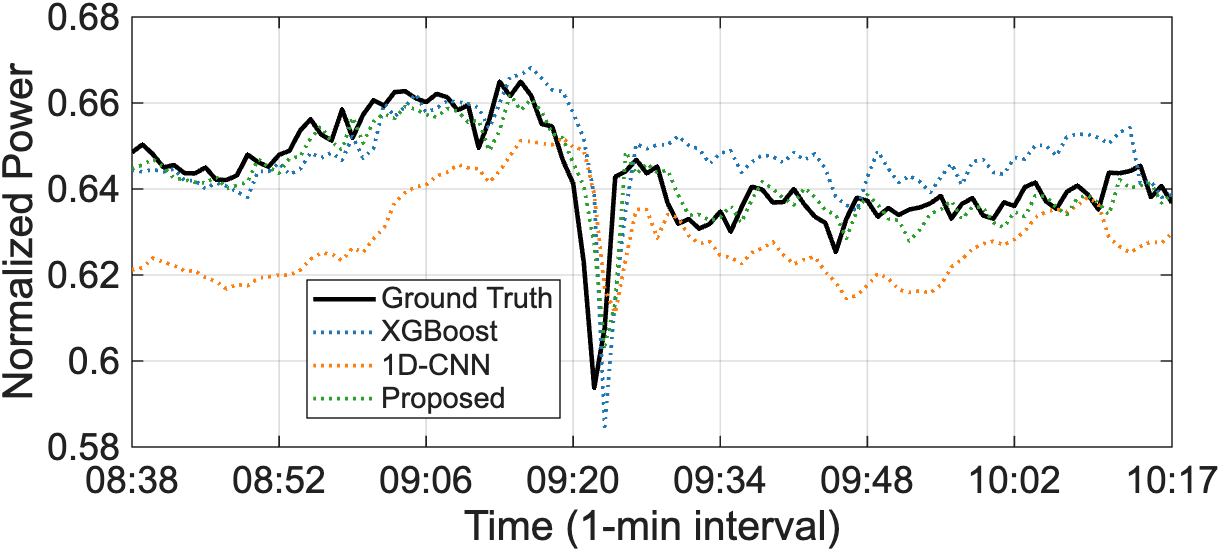}%
    \label{fig:Figure_2}}
    \\[4pt]
  \subfloat[]{%
    \includegraphics[width=0.40\textwidth]{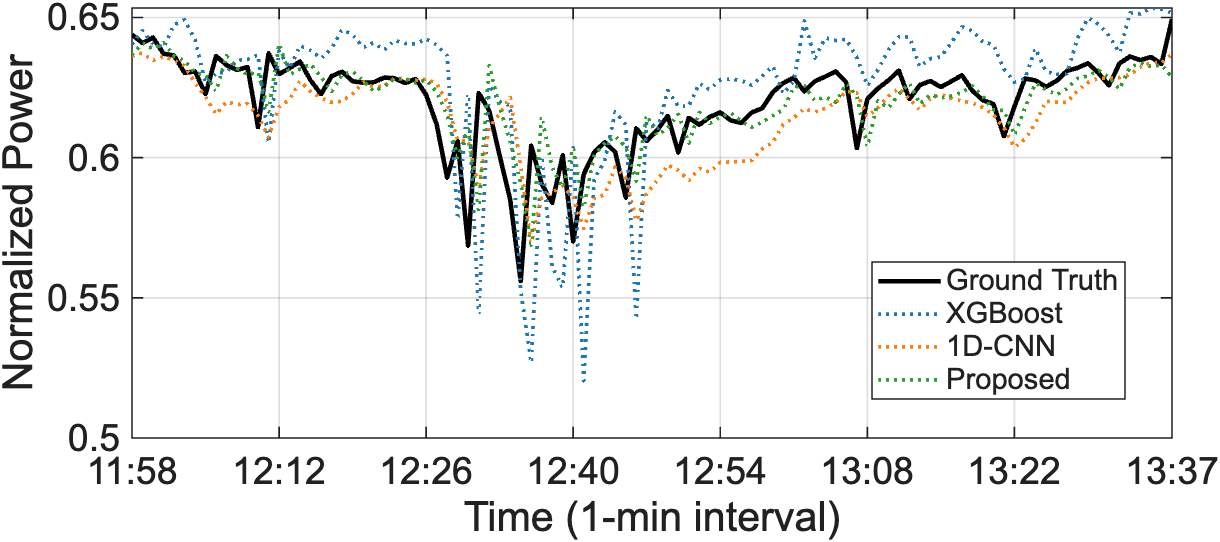}%
    \label{fig:Figure_3}}
    \\[3pt]
  \subfloat[]{%
    \includegraphics[width=0.40\textwidth]{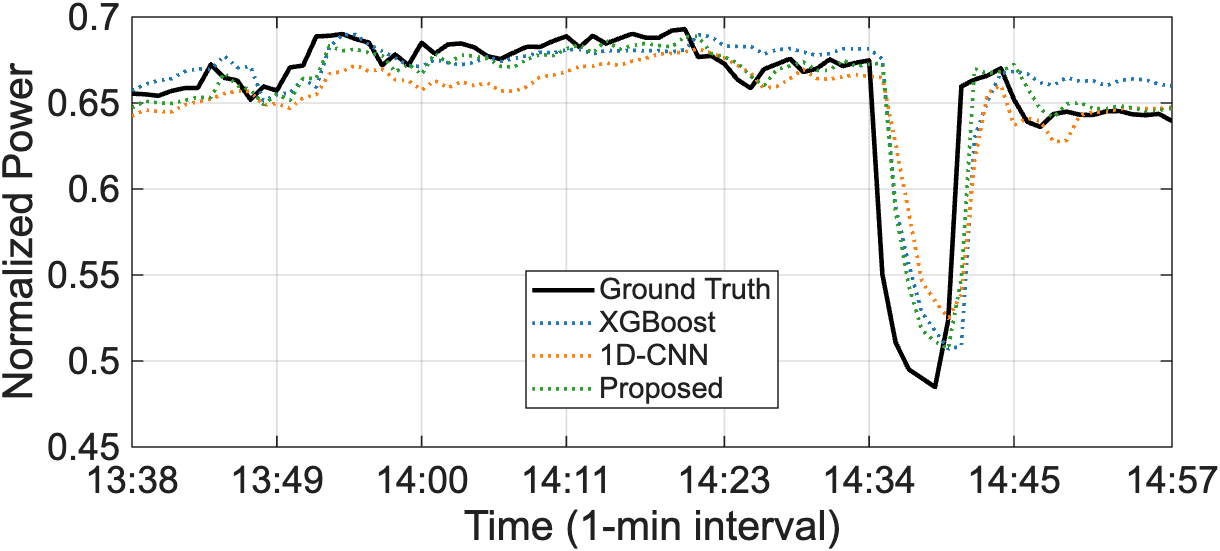}%
    \label{fig:Figure_4}}
  \caption{Power forecasting results over a future forecasting horizon involving multiple regime shifts. (a) Overview and (b)–(d) Enlarged views of the three marked segments.}
  \label{fig:ensemble_best_pair}
\end{figure}

\subsection{Performance Evaluation of Proposed Method}
The forecasting accuracy of the proposed ensemble learning method is evaluated against four ML methods, while its adaptivity across operating regimes is verified relative to its two submodels.

Table~\ref{tab:one_step_error_comparison} compares the forecasting errors of different algorithms. The proposed method achieves the lowest errors among the five ML methods. Compared with LSTM and SVR, the reductions are 68.7\% and 72.6\% in NRMSE, and 82.6\% and 84.0\% in NMAE, respectively. Relative to the strongest baseline, 1D-CNN, which achieves an NRMSE of 2.45\% and an NMAE of 1.85\%, the proposed method still leads to relative improvements of 66.1\% and 80.2\%, respectively.

The forecasting results over a representative horizon involving multiple regime shifts are illustrated in Fig.~\ref{fig:ensemble_best_pair}, where neither XGBoost nor 1D-CNN alone consistently tracks the ground truth in the selected test window. However, the ensemble output exhibits markedly better alignment with the ground truth. This indicates that the two submodels, despite their individual errors, provide complementary predictive information that can be effectively exploited through ensemble learning. These results demonstrate that the proposed method substantially enhances forecasting accuracy over either individual submodel.

\begin{figure}[!t]{
  \centering
  \includegraphics[width=0.9\linewidth]{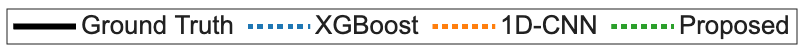} 
  \vspace{-5pt}
  \subfloat[]{%
    \includegraphics[width=0.56\linewidth]{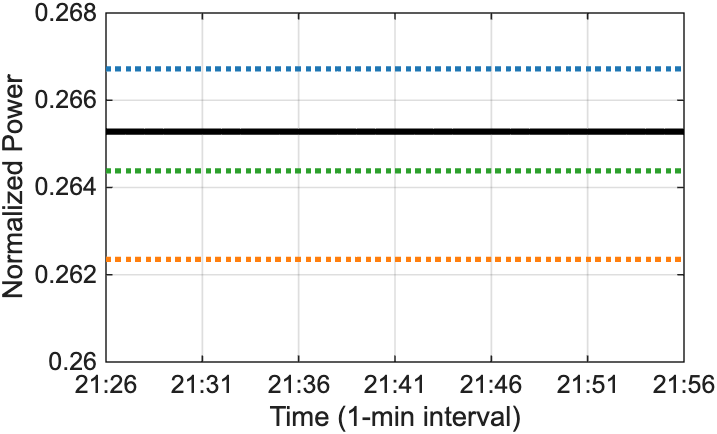}
    \label{fig:idle}}
  \subfloat[]{%
    \includegraphics[width=0.40\linewidth]{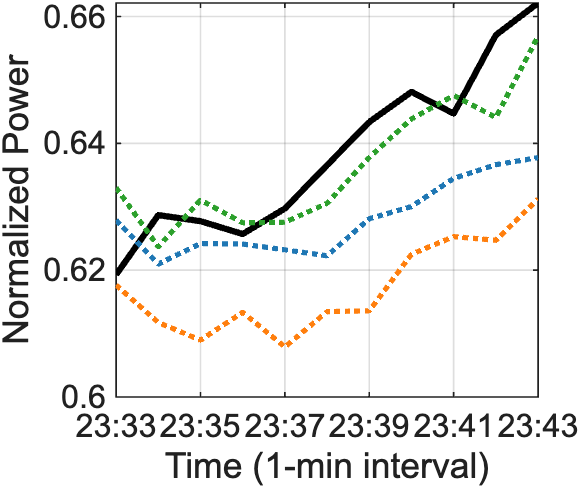}
    \label{fig:ramp-up}}
  \hfill
  \subfloat[]{%
    \includegraphics[width=0.57\linewidth]{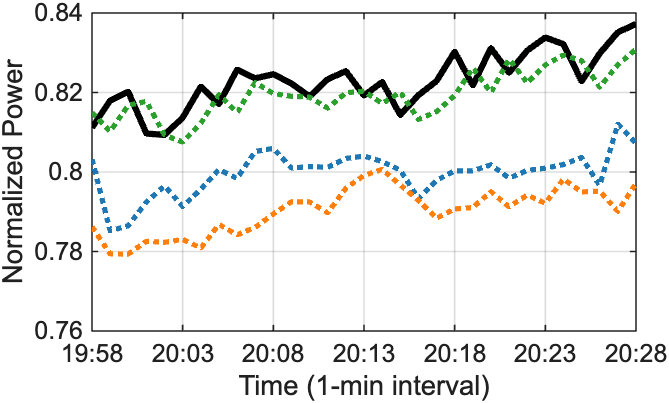}
    \label{fig:steady}}
  \subfloat[]{%
    \includegraphics[width=0.40\linewidth]{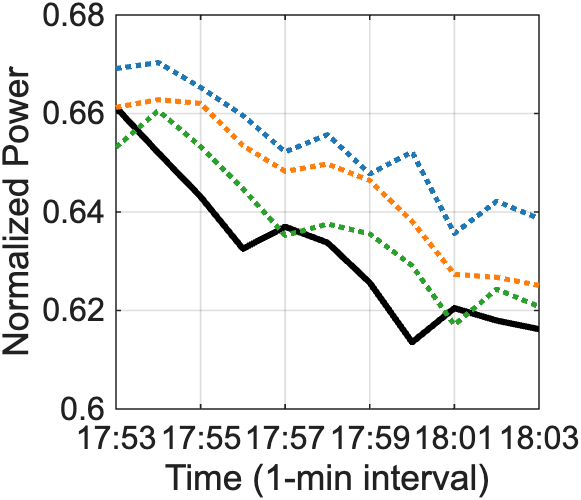}
    \label{fig:ramp-down}}
  \caption{Prediction performance under four different operating regimes: (a) idle, (b) ramp-up, (c) high, and (d) ramp-down demands.} 
    \label{fig:regimes}}
\vspace{-8pt}
\end{figure}

Fig.~\ref{fig:regimes} compares the forecasting performance of XGBoost, 1D-CNN, and the proposed ensemble under four representative operating regimes: idle (no active training jobs), ramp-up (job launch with increasing workload intensity), high (sustained high workload intensity from concurrent training jobs), and ramp-down (job completion with subsiding workload intensity). Distinct regime-dependent behaviors are observed for the two submodels. In the idle regime, XGBoost overestimates while 1D-CNN underestimates the near-constant ground truth. In both the ramp-up and ramp-down regimes, the submodels exhibit systematic biases throughout the transitions. In the high regime, both submodels consistently underestimate the fluctuating trajectory. Across all regimes, the proposed ensemble follows the ground truth more closely, confirming that it adaptively exploits the complementary strengths of the two submodels.

\begin{figure*}
  \centering
  \vspace{5pt}
  \subfloat[]{%
    \includegraphics[width=0.35\linewidth]{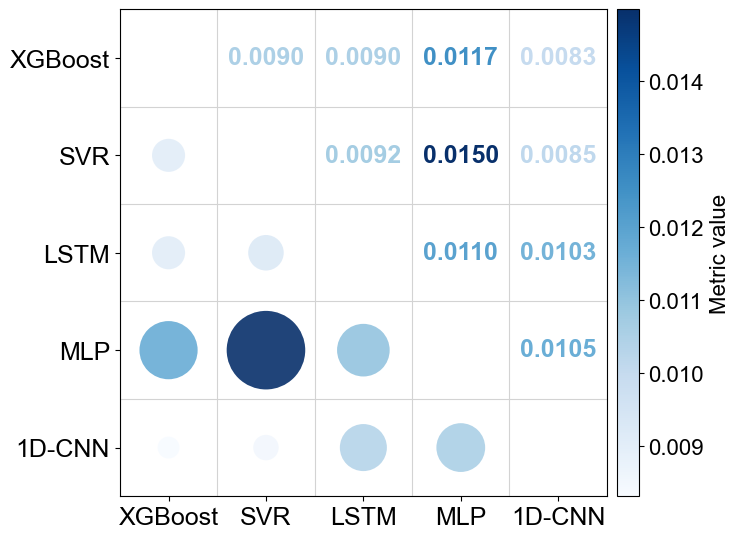}%
    \label{fig:submodels-nrmse}}\qquad \qquad
  \subfloat[]{%
    \includegraphics[width=0.35\linewidth]{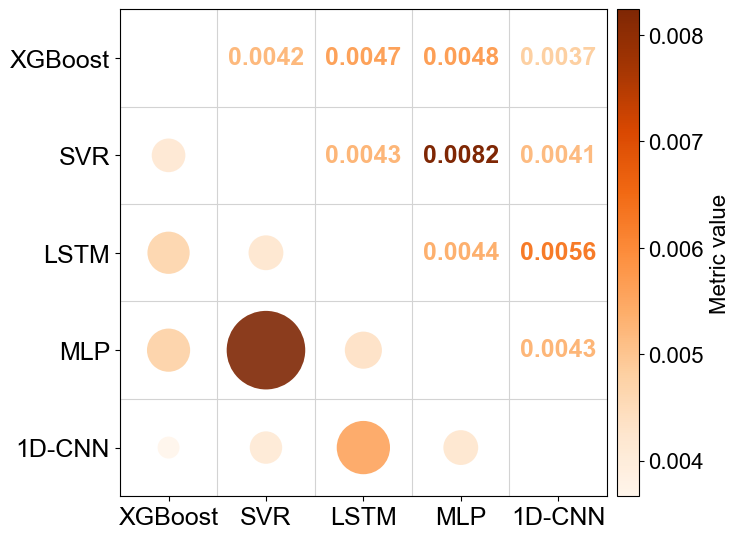}%
    \label{fig:submodels-nmae}}
  \caption{Forecasting errors of two-model ensembles for one-step-ahead prediction. (a) NRMSE. (b) NMAE.}
  \label{fig:ensemble_performance}
\end{figure*}

\subsection{Comparative Analysis of Alternative Submodel Pairs}
The model complementarity and forecasting errors are investigated for all pairwise ensembles of the five recent ML models, XGBoost, 1D-CNN, LSTM, MLP, and SVR. 
To quantify the balance in the Talagrand distribution, the standard deviation of the three category frequencies is calculated as \cite{li2023adaptive}
\begin{equation}
\sigma_{\mathrm{RH}}
=
\sqrt{
\frac{1}{3}
\sum_{k=1}^{3}
\left(F_k-\frac{1}{3}\right)^2
}
\end{equation}
where $F_k = \frac{N_k}{N}$, and $k=1,2,3$; $N_k$ and $N$ denote the number of samples falling into the $k$th category and the total number of test samples.
A smaller $\sigma_{\mathrm{RH}}$ indicates a Talagrand distribution closer to the uniform distribution, reflecting stronger complementarity between the two submodels.

\begin{table}[!t]
\captionsetup{font=small,justification=centering}
\caption{Comparison of Talagrand Distributions for Submodel Pairs}
\centering
\label{tab:talagrand_pairs}
\begin{tabular}{|l|c|c|c|c|}
\hline
\multirow{2}{*}{Ensemble Pair} & \multicolumn{3}{c|}{Frequency of Each Category} & \multirow{2}{*}{$\sigma_{\mathrm{RH}}$} \\
\cline{2-4}
& 1 & 2 & 3 & \\
\hline
\textbf{XGBoost--1D-CNN} & \textbf{0.3307} & \textbf{0.3145} & \textbf{0.3547} & \textbf{0.0165} \\
\hline
LSTM--MLP       & 0.2847 & 0.3419 & 0.3734 & 0.0367 \\
\hline
XGBoost--SVR    & 0.3334 & 0.2882 & 0.3784 & 0.0368 \\
\hline
SVR--LSTM       & 0.3744 & 0.2679 & 0.3577 & 0.0468 \\
\hline
XGBoost--LSTM   & 0.4265 & 0.2949 & 0.2786 & 0.0662 \\
\hline
LSTM--1D-CNN    & 0.4212 & 0.1954 & 0.3834 & 0.0988 \\
\hline
XGBoost--MLP    & 0.1949 & 0.4597 & 0.3453 & 0.1084 \\
\hline
SVR--MLP        & 0.2291 & 0.2602 & 0.5107 & 0.1261 \\
\hline
1D-CNN--MLP     & 0.2581 & 0.2233 & 0.5186 & 0.1318 \\
\hline
SVR--1D-CNN     & 0.3492 & 0.1464 & 0.5044 & 0.1466 \\
\hline
\end{tabular}%
\vspace{-7pt}
\end{table}

Table~\ref{tab:talagrand_pairs} lists the Talagrand distributions of all ensemble pairs. The XGBoost--1D-CNN pair yields the smallest Talagrand deviation of 0.0165, with category frequencies closest to a uniform distribution, indicating the strongest complementarity among the examined pairs. In contrast, SVR--1D-CNN, 1D-CNN--MLP, and SVR--MLP show substantially larger deviations, reflecting lower complementarity. The XGBoost--1D-CNN pair is thus adopted in the proposed ensemble algorithm.

Fig.~\ref{fig:ensemble_performance} compares the forecasting errors of all ensemble pairs. The XGBoost--1D-CNN ensemble pair achieves the best performance, yielding the lowest NRMSE and NMAE among all the combinations. By contrast, the SVR--MLP pair produces the largest errors and thus exhibits the weakest forecasting performance. In addition, other pairings, such as MLP--1D-CNN and SVR--1D-CNN, remain competitive but are still inferior to the XGBoost--1D-CNN ensemble.
These observations are consistent with the Talagrand distribution analysis, which shows that the adopted XGBoost--1D-CNN pairing exhibits stronger complementarity than the alternatives.

\section{Conclusion}
This paper proposes an adaptive weighting ensemble learning method for short-term load forecasting in AI data centers. The proposed algorithm dynamically adjusts ensemble weights according to the incremental load dynamics and historical submodel performance, by fully exploiting the complementary strengths of top-performing ML submodels, XGBoost and 1D-CNN. Case studies show that the proposed method outperforms the individual submodels and remains robust under different operating regimes. The results demonstrate its effectiveness in capturing the bursty non-stationary load behavior of AI data centers, thereby highlighting its potential for grid-interactive coordination and demand response.

\bibliographystyle{IEEEtran}
\bibliography{IEEEabrv,Citation}
\let\mybibitem\bibitem

\end{document}